\documentstyle[12pt]{elsart}
\begin{document}
\begin{frontmatter}
\title{\hfill MZ-TH/99-45\\[5mm]
$\Lambda_b$ and $\Lambda_c$ baryon decays at finite values of
heavy quark masses}
\author{M.\ A.\ Ivanov}
\address{Bogoliubov Laboratory of Theoretical Physics, Joint Institute
for Nuclear Research, 141980 Dubna, Russia}
\author{J.\ G.\ K\"{o}rner}
\address{Johannes Gutenberg-Universit\"{a}t, Institut f\"{u}r Physik,
D-55099 Mainz, Germany}
\author{V.\ E.\ Lyubovitskij}
\address{Bogoliubov Laboratory of Theoretical Physics, Joint Institute
for Nuclear Research, 141980 Dubna, Russia
and Department of Physics, Tomsk State University,
634050 Tomsk, Russia}
\author{A.\ G.\ Rusetsky}
\address{Institute for Theoretical Physics,
University of Bern, Sidlerstrasse 5, CH-3012, Bern, Switzerland and
Bogoliubov Laboratory of Theoretical Physics, Joint Institute
for Nuclear Research, 141980 Dubna, Russia and HEPI, Tbilisi State
University, 380086 Tbilisi, Georgia}

\maketitle

\begin{abstract}
Semileptonic decays of $\Lambda_b$ and $\Lambda_c$ baryons are
studied within the Relativistic Three-Quark Model using finite heavy
quark mass values. Employing the same parameters as have been used
previously for the description of exclusive decays of heavy baryons
in the heavy quark limit we calculate the six form factors of the
process and the corresponding decay rates. Our calculation shows that the
``finite mass'' corrections are important in heavy-to-light
transitions and are not negligible in heavy-to-heavy transitions.
\end{abstract}
\end{frontmatter}

\noindent
{\it Pacs:} 11.15.Me, 11.30.Hv, 12.38.Lg, 12.39.Ki, 14.20.Lq, 14.20.Mr


During the last decade heavy baryon transitions (semileptonic,
nonleptonic, strong and electromagnetic) have been studied in detail
within the Heavy Quark Effective Theory employing QCD sum rule methods or
nonrelativistic and relativistic quark models, etc. (see, for example,
the reviews in \cite{KKP,MN} and the papers
\cite{IW,HKKG,MRR,EIKL,H,KM,Yan,Cheng,HDH,ILKK,IKLR1,IKLR2,IKLR3,TOK,DFNR}). 
The mass spectrum of heavy baryons as well as their exclusive and inclusive 
decays have been described succesfully in these approaches incorporating
the ideas of QCD. 

In the papers \cite{ILKK,IKLR1,IKLR2,EIL,EI,AIKL,ILL} we proposed and
developed a QCD motivated Relativistic Three-Quark Model (RTQM),
which can be viewed as an effective quantum field approach based on an 
interaction Lagrangian of light and heavy baryons interacting with their 
constituent quarks. The coupling strength of the baryons interacting with 
the three constituent quarks
are determined by the compositeness condition $Z_H=0$  \cite{EI,SWH}
where $Z_H$ is the wave function renormalization constant of the hadron.
The compositeness condition enables one to unambiguously and consistently
relate theories with quark and hadron degrees of freedom
to the effective Lagrangian approaches formulated in terms of hadron
variables only (as, for example, Chiral Perturbation Theory \cite{WGL}
and its covariant extension to the baryon sector \cite{BL}).
Our strategy is as follows. We start with an effective interaction
Lagrangian written down in terms of quark and hadron variables. 
Then, by using Feynman rules, the $S$-matrix elements describing 
hadron-hadron interactions are given in terms of a set of quark diagrams. 
The compositeness condition enables one to avoid a double counting of quark 
and hadron degrees of freedom. The RTQM model contains only a few model
parameters: the masses of the light and heavy quarks, and certain scale 
parameters that define the size of the distribution of the constituent 
quarks inside the hadron. The RTQM approach has been previously used to 
compute the exclusive semileptonic, nonleptonic, strong and electromagnetic
decays of charm and bottom baryons \cite{ILKK,IKLR1,IKLR2} in the heavy
quark limit  $m_Q\to\infty$ always employing the same set of model parameters.

The main objective of this letter is to extend our approach to the
study of heavy baryon transitions at finite values of the heavy quark
mass without using an explicit $1/m_Q$ expansion. In the following we confine
ourself to the dominant exclusive semileptonic transitions of $\Lambda_b$ and 
$\Lambda_c$ baryons: $\Lambda_b\to\Lambda_c e^- \bar\nu_e$ and
$\Lambda_c\to\Lambda_s e^+ \nu_e$. We compare our results to a similar 
finite mass calculation done in \cite{DFNR}, where the QCD sum rule approach 
was used. In agreement with \cite{DFNR} we find that finite quark mass 
corrections to the form factors and the rates of semileptonic transitions 
are important for heavy-to-light $c \rightarrow s$  transitions and not
negligible for heavy-to-heavy $b \rightarrow c$ transitions.

We proceed as follows. First we briefly explain the basic ideas of
the Relativistic Three-Quark Model (RTQM) and describe our
calculational techniques for the finite heavy quark case.
We then compute form factors and rates for the decays
$\Lambda_b\to\Lambda_c e^- \bar\nu_e$ and
$\Lambda_c\to\Lambda_s e^+ \nu_e$. For reasons of comparison numerical
results are provided both for the finite quark mass case and the infinite
quark mass case. We finally compare our results with the results of recent
calculations within the QCD sum rule method \cite{DFNR}.

We start with a brief description of our approach,
the Relativistic Three-Quark Model (RTQM). A detailed description of
the RTQM can be found in Refs. \cite{ILKK,IKLR2,AIKL,ILL}.
In the RTQM approach baryons are described as bound states of
consitituent quarks. The coupling of the baryons with their constituents
is defined through an effective relativistic interaction Lagrangian which
contains the usual three-quark currents with the quantum numbers of 
the heavy baryons \cite{ILKK}. For example,
the Lagrangian of the $\Lambda$-type baryons
($\Lambda_b$, $\Lambda_c$ and $\Lambda_s$) coupling to three quarks
is taken as
\begin{eqnarray}\label{LHB_int}
{\cal L}_{\Lambda_q}^{\rm int}(x)&=&g_{B_q}
\bar\Lambda_q(x)\Gamma_1 {q}^a(x)
\int\hspace*{-0.1cm} d\xi_1\hspace*{-0.1cm}
\int\hspace*{-0.1cm} d\xi_2
F_{B_q}(\xi_1^2+\xi_2^2)\\[2mm] \label{int_bar}
&\times&u^b(x+3\xi_1-\sqrt{3}\xi_2)C\Gamma_2
d^c(x+3\xi_1+\sqrt{3}\xi_2)\varepsilon^{abc}+{\rm h.c.}\nonumber\\[2mm]
F_{B_q}(\xi_1^2+\xi_2^2)&=&
\int\hspace*{-0.1cm}\frac{d^4k_1}{(2\pi)^4} \hspace*{-0.1cm}
\int\hspace*{-0.1cm}\frac{d^4k_2}{(2\pi)^4} \hspace*{0.1cm}
e^{ik_1\xi_1+ik_2\xi_2}\tilde F_{B_{\cal Q}}(k_1^2+k_2^2)
\nonumber
\end{eqnarray}
where the $q=b, c, s$ stands for the quark spinor and $\Gamma_i$ are spinor
matrices which define the quantum numbers of the relevant three-quark
currents. $C=\gamma^0\gamma^2$ is the charge conjugation matrix and
$g_{B_q}$ is the coupling constant which is determined by
the compositeness condition \cite{IKLR2,EI,SWH}. The compositeness
condition implies that the renormalization constant of the hadron wave
function is set equal to zero: $Z_{B_q} = 1 - g_{B_q}^2
\Sigma^\prime_{B_q}(M_{B_q}) = 0$ where
$\Sigma^\prime_{B_q}$ is the derivative of the baryon mass operator and
$M_{B_q}$ is the baryon mass. 

It is well known that the
form of the three-quark current for baryons is not unique. In fact, one
can write down different interpolating currents for a given baryon (even in 
the absence of derivative interactions) that have  the correct quantum numbers 
of the given baryon. For finite mass $\Lambda$-type baryons there are three 
possibilities for the choice of the three-quark current without derivatives. 
In papers \cite{EIL,IKLPR} we showed that baryon observables are only weakly 
dependent on the choice of the three-quark currents. 
In this paper we consider the simplest set of baryon currents:
the pseudoscalar currents for the heavy baryons $\Lambda_b$ and $\Lambda_c$
($\Gamma_1\otimes C\Gamma_2=I\otimes C\gamma_5$) \cite{ILKK} and a
$SU(3)$ symmetric (tensor) current for the $\Lambda_s$ hyperon 
($\Gamma_1\otimes C\Gamma_2=I\otimes C\gamma_5+\gamma_5\otimes C$) \cite{EIL}. 

In Eq. (\ref{LHB_int}) we have introduced a baryon-three-quark vertex
form factor given by $\tilde F_{B_q}(k_1^2+k_2^2)$.
For simplicity, we factorize out the $q=b, c$ or $s$ quark in the
$\Lambda$-type baryons by placing them at the center of the baryon.
This
means that we take the masses of the $b$, $c$ and $s$ quarks to be much
larger than the masses of $u$ and $d$ quarks.
Any choice of vertex function $F_{B_q}$ is appropiate as long
as it falls off sufficiently fast in the ultraviolet region to
render the Feynman diagrams ultraviolet finite.
In principle, their functional forms would be calculable from the solutions
of the Bethe-Salpeter equations for the baryon bound states \cite{IKLR3}
which is, however, an untractable problem at present.
In our previous analysis \cite{AIKL} we found that, using various
forms for the vertex function, the hadron observables
are insensitive to the exact details of
the functional form of the hadron-quark vertex form factor.
We will use this observation as a guiding principle and choose simple
Gaussian forms for the vertices $F_{B_q}$.
Their Fourier transform reads \cite{ILKK,IKLR1,IKLR2,ILL}
\begin{equation}\label{BS}
\tilde F_{B_q}(k_1^2+k_2^2)=
\exp\left(\frac{k_1^2+k_2^2}{\Lambda_{B_q}^2}\right)
\end{equation}
where $\Lambda_{B_q}$ is a scale parameter defining the distribution
of the $u$ and $d$ quarks in $\Lambda$-type baryon.
For the light and heavy quark propagators with constituent masses $M$
we shall use the standard form of the free fermion propagator
\begin{equation}\label{light}
S_M(k)=\frac{1}{M-\not\! k}
\end{equation}
where $M=m$ for the $u$ or $d$ quarks, $M=m_s$ for the strange
quark, $M=m_c$ for the charm quark and $M=m_b$ for the bottom quark.

Next we specify our model parameters. In order to be able to compare the two
calculations with finite quark masses on the one hand and infinite quark
masses \cite{IKLR2} on the other hand
we use the same set of model parameters in both calculations:
i)$\Lambda_{B}=\Lambda_{B_u}=\Lambda_{B_d}=\Lambda_{B_s}=1$ GeV is the
common  scale parameter defining the distribution of quarks in
light baryons \cite{ILL};
ii) in the heavy flavour sector the scale parameters $\Lambda_{B_c}$ and
$\Lambda_{B_b}$ are chosen to be the same
$\Lambda_{B_Q}=\Lambda_{B_c}=\Lambda_{B_b}=1.8$ GeV in order to provide
the correct normalization of the baryonic Isgur-Wise function
in the heavy quark limit \cite{IKLR2};
iii) the values of the constituent quark masses are fixed from the
analysis of magnetic moments and charge radii of light baryons
\cite{IKLR2,ILL}: $m=420$ MeV and $m_s=570$ MeV.
Therefore, we have only two free parameters: $m_c$ and $m_b$, the masses of
the charm and bottom quark. Their values are fixed according to \cite{IKLR2}
\begin{eqnarray}\label{condition}
\bar\Lambda=M_{\Lambda_b}-m_b=M_{\Lambda_c}-m_c=600 \,\,\, \mbox{MeV}
\end{eqnarray}
In this way one can meaningfully compare the results of the two
calculations for finite and infinite quark mass values.
Using Eq. (\ref{condition}) with the experimental values
for the $\Lambda_b$ and $\Lambda_c$ baryon masses
$m_{\Lambda_b}=5.64$ GeV and $m_{\Lambda_c}=2.285$ GeV \cite{PDG} we find
$m_b=5.04$ GeV and $m_c=1.685$ GeV.
The results obtained in this paper will be compared to the results 
of \cite{IKLR2} done in the heavy quark limit $(m_Q\to\infty)$.
In both cases the difference between the masses of the heavy baryons
and the masses of the heavy quarks is fixed at $\bar\Lambda = 600$ MeV. 
The two calculations differ only in the choice of the heavy quark
propagators: in the finite mass scheme we use the usual free propagator
(without any $1/m_Q$ expansion) and in the infinite mass scheme we use
the usual leading HQET propagator \cite{ILKK,IKLR2}
\begin{eqnarray}\label{heavy_propagator}
S_v(k,\bar\Lambda)=&-\frac{(1+\not\! v)}
{2(v\cdot k + \bar\Lambda)}
\end{eqnarray}
where the four-velocity of the heavy quark is denoted by $v$ as usual. 

The semileptonic decays of the $\Lambda_b$ and $\Lambda_c$ baryons are
described by the triangle two-loop diagram shown Fig.1.
Correspondingly one has
\begin{eqnarray}\label{vertex}
\hspace*{-.75cm}M_\mu(\Lambda_{q_1}\to \Lambda_{q_2})=
\bar u(p_2) \Lambda_\mu(p_1,p_2) u(p_1)=
\frac{g_{B_{q_1}}}{8\pi^2}\frac{g_{B_{q_2}}}{8\pi^2}
\bar u(p_2) \Gamma_\mu(p_1,p_2)u(p_1)
\end{eqnarray}
for the matrix element of the transition.
The weak vertex function $\Gamma_\mu(p_1,p_2)$ takes the form
\begin{eqnarray}\label{loop}
\hspace*{-.75cm}\Gamma_\mu(p_1,p_2)&=&
\int\hspace*{-0.1cm}\frac{d^4k_1}{\pi^2i} \hspace*{-0.1cm}
\int\hspace*{-0.1cm}\frac{d^4k_2}{4\pi^2i} \hspace*{0.1cm}
\tilde F_{B_{q_1}}(12[k_1^2+k_1k_2+k_2^2])
\tilde F_{B_{q_2}}(12[k_1^2+k_1k_2+k_2^2])\nonumber\\
\hspace*{-.75cm}&\times&
\Gamma_1^f S_{q_2}(k_1+p_2)O_\mu S_{q_1}(k_1+p_1)\Gamma_1^i
{\rm Tr}[\Gamma^f_2 S_{q_4}(k_2) \Gamma^i_2 S_{q_3}(k_1+k_2)]
\end{eqnarray}
where $\Gamma_{1(2)}^i$ and $\Gamma_{1(2)}^f$
are the Dirac matrices of the initial and the final baryons, respectively,
and $O_\mu=\gamma_\mu(1+\gamma_5)$.
The integral (\ref{loop}) is calculated in the Euclidean
region both for internal and external momenta. The final results are
obtained by analytic continuation of the external momenta to the physical
region after the internal momenta have been integrated out.
In order to keep the calculation as general as possible 
we shall retain a general form for the
vertex function $\tilde F_{B_{q}}$ in our analytical integrations.
Only at the very end of the calculation when we do the numerical 
evaluation the Gaussian form (\ref{BS}) will be substituted
for the vertex function. As mentioned before the difference to the earlier
calculation in Ref. \cite{IKLR2} lies in the use of the full quark propagator,
whereas in \cite{IKLR2} we have used the leading HQET propagator.
The integration techniques  used in \cite{IKLR2} can be easily extended
to the case of finite heavy quark masses.
 
As an illustration of our calculational procedure we evaluate integral
(\ref{loop}) for equal values of the heavy and light scale parameters
$\Lambda_{B_{q_1}}=\Lambda_{B_{q_2}}=\Lambda_{B_Q}$.
First of all, we express all dimensional parameters entering the
two-loop integral (\ref{loop}) in units of $\Lambda_{B_Q}$. We then have
\begin{eqnarray}
\hspace*{-.75cm}
\Gamma_\mu(p_1,p_2)&=&\int\limits_0^\infty \hspace*{-0.1cm}ds
\tilde F_B^L(s) \int\limits_0^\infty \hspace*{-0.1cm}d^4\alpha \,\,\,
e^{-m_3^2\alpha_3-m_4^2\alpha_4-(m_1^2-p_1^2)\alpha_1-(m_2^2-p_2^2)\alpha_2}\\
\hspace*{-.75cm}&\times&\Gamma_1^f \biggl(m_2 + \not\! p_2 +
\frac{\not\! \partial_1}{2}\biggr)
O_\mu \biggl(m_1 + \not\! p_1 + \frac{\not\! \partial_1}{2}
\biggr)
\Gamma_1^i\nonumber\\
\hspace*{-.75cm}&\times& {\rm Tr}\biggl[\Gamma^f_2
\biggl(m_4 + \frac{\not\! \partial_2}{2}\biggr)
\Gamma_2^i\biggl(m_3+\frac{\not\! \partial_1}{2}+\frac{\not\! \partial_2}{2}
\biggr)\biggr]
\int\hspace*{-0.1cm}\frac{d^4k_1}{\pi^2i} \hspace*{-0.1cm}
\int\hspace*{-0.1cm}\frac{d^4k_2}{4\pi^2i} \hspace*{0.1cm}
e^{k{\cal A}k+2kB}\nonumber
\end{eqnarray}
where
$\not\! \partial_i = \gamma_\mu \cdot \partial / \partial B_i^\mu$, 
$\tilde F_B^L(s)$ is the Laplace transform of 
$\tilde F_{B_Q}^2(12[k_1^2+k_1k_2+k_2^2])$ and the matrices 
${\cal A}$ and $B$ are defined by
\[{\cal A}_{ij}=\left(
\begin{array}{ll}
\mbox{$12s+\alpha_1+\alpha_2+\alpha_3$}&\hspace*{.5cm}\mbox{$6s+\alpha_3$}\\
\mbox{$6s+\alpha_3$}&\hspace*{.5cm}\mbox{$12s+\alpha_3+\alpha_4$}
\end{array}
\right) \]
\[B_{i}=\left(\begin{array}{l}
 \mbox{$\alpha_1p_1+\alpha_2p_2$} \\
\mbox{$\,\,\,\,\,\,\,\,\,\,\,\,\,\,0$}
\end{array}
\right)\]

The integration over $k_1$, $k_2$ and the variable $s$ results in
\begin{eqnarray}
& &\Gamma_\mu(p_1,p_2)=\int\limits_0^\infty \hspace*{-0.1cm}d^4\alpha
\biggl\{\frac{\tilde F_{B_Q}^2(-12w)}{|A|^2}
\Gamma_1^f (m_2 + \not\! p_2 - \not\! B_1 A^{-1}_{11})O_\mu\\[2mm]
&\times&(m_1 + \not\! p_1 - \not\! B_1 A^{-1}_{11}) \Gamma_1^i
{\rm Tr}[\Gamma_2^f(m_4 - \not\! B_1 A^{-1}_{12})
\Gamma_2^i(m_3 - \not\! B_1 (A^{-1}_{11}+A^{-1}_{12}))]\nonumber\\[2mm]
&-&\int\limits_0^\infty dt
\frac{\tilde F_{B_Q}^2(-12[w+t])}{2|A|^2}\biggl[(A^{-1}_{12}+A^{-1}_{22})
{\rm Tr}[\Gamma_2^f\gamma^\beta\Gamma_2^i\gamma_\beta]
\Gamma_1^f(m_2 +\not\! p_2-\not\! B_1 A^{-1}_{11})\nonumber\\[2mm]
&\times&O_\mu(m_1 + \not\! p_1-\not\! B_1A^{-1}_{11})\Gamma_1^i
+\Gamma_1^f[\gamma^\alpha O_\mu (m_1 + \not\! p_1 - \not\! B_1 A^{-1}_{11})
\nonumber\\[2mm]
&+&(m_2 + \not\! p_2 - \not\! B_1 A^{-1}_{11}) O_\mu \gamma^\alpha]\Gamma_1^i
\biggl(A^{-1}_{12}{\rm Tr}[\Gamma_2^f\gamma_\alpha \Gamma_2^i
(m_3- \not\! B_1 (A^{-1}_{11}+A^{-1}_{12})]\nonumber\\[2mm]
&+&(A^{-1}_{11}+A^{-1}_{12})
{\rm Tr}[\Gamma_2^f(m_4- \not\! B_1 A^{-1}_{12}) \Gamma_2^i\gamma_\alpha]
\biggr)\nonumber\\[2mm]
&+&\Gamma_1^f\gamma^\alpha O_\mu\gamma_\alpha \Gamma_1^i A^{-1}_{11}
{\rm Tr}[\Gamma_2^f(m_4 - \not\! B_1 A^{-1}_{12})
\Gamma_2^i(m_3 - \not\! B_1 (A^{-1}_{11}+A^{-1}_{12}))]\biggr]
\nonumber\\[5mm]
&+&\int\limits_0^\infty dtt\frac{\tilde F_{B_Q}^2(-12[w+t])}{4|A|^2}
\biggl[\Gamma_1^f\gamma^\alpha O_\mu\gamma_\alpha \Gamma_1^i
A^{-1}_{11}(A^{-1}_{12}+A^{-1}_{22}){\rm Tr}[\Gamma_2^f\gamma^\beta
\Gamma_2^i \gamma_\beta]\nonumber\\[2mm]
&+&\Gamma_1^f \gamma^\alpha O_\mu\gamma^\beta\Gamma_1^i
A^{-1}_{12} (A^{-1}_{11}+A^{-1}_{12}){\rm Tr}[\Gamma_2^f\gamma_\alpha
\Gamma_2^i\gamma_\beta+\Gamma_2^f\gamma_\beta\Gamma_2^i\gamma_\alpha]\biggr]
\biggr\}\nonumber
\end{eqnarray}
where $M_1$ and $M_2$ are the masses of initial and final baryons,
respectively, and
$$w=\sum\limits_{i=1}^4 m_i^2\alpha_i - M_1^2\alpha_1
- M_2^2\alpha_2 + A^{-1}_{11}(p_1\alpha_1+p_2\alpha_2)^2,$$
\[A^{-1}_{ij}=\frac{1}{|A|}\left(
\begin{array}{ll}
 \mbox{$1+\alpha_3+\alpha_4$} & \hspace*{.2cm} \mbox{$-(1/2+\alpha_3)$}\\
 \mbox{$-(1/2+\alpha_3)$} & \hspace*{.2cm} \mbox{$1+\alpha_1 +
 \alpha_2 + \alpha_3$}
\end{array}
\right) \]

The vertex function $\Lambda_\mu(p_1,p_2)$ is as usual decomposed into a set
of six relativistic form factors which are functions of the 
momentum transfer squared $t=q^2$. We shall present our results
in terms of two alternative sets of heavy baryon weak form factors 
\cite{KKP,IW,EIKL}.
The two sets of form factors are defined by the covariant expansions 
\begin{eqnarray}
\Lambda_\mu(p_1,p_2)=\gamma_\mu (F_1^V+F_1^A\gamma_5)
+ i\sigma_{\mu\nu}q^\nu (F_2^V+F_2^A\gamma_5)
+ q^\mu (F_3^V+F_3^A\gamma_5)
\end{eqnarray}
and
\begin{eqnarray}
\Lambda_\mu(p_1,p_2)=
\gamma_\mu (G_1^V+G_1^A\gamma_5)
+ v_\mu (G_2^V+G_2^A\gamma_5)
+ v_\mu^\prime (G_3^V+ G_3^A\gamma_5)
\end{eqnarray}
where $v_\mu=p_\mu/M_1$ and $v_\mu^\prime=p_\mu^\prime/M_2$ are the
four-velocities of the initial and final baryon, respectively.
The relation between the two sets of heavy baryon form factors can be
easily worked out \cite{KKP,EIKL}.
In the heavy quark limit \cite{IW} one has
\begin{eqnarray}
F_1^V=F_1^A=G_1^V=G_1^A, \hspace*{.5cm}
F_2^{V(A)}=F_3^{V(A)}=G_2^{V(A)}=G_3^{V(A)}=0
\end{eqnarray}
for the heavy-to-heavy transition $\Lambda_b\to\Lambda_c e^-\bar\nu_e$.
For heavy-to-light transitions as in $\Lambda_c\to\Lambda_s e^+\nu_e$
transition one has
\begin{eqnarray}
G_2^V=G_2^A=G_1^A-G_1^V, \hspace*{.15cm}
G_3^V=G_3^A=0, \hspace*{.15cm}
F_1^V=F_1^A, \hspace*{.15cm} F_2^{V(A)}=F_3^{V(A)}
\end{eqnarray}
in the heavy ($c$-quark) limit \cite{HKKG,H}. We shall not write down rate 
and asymmetry formulae in terms of these form factors since these have been 
worked out in great detail in Refs. \cite{KKP,EIKL,KM,ILKK}.

We now present our numerical results for the exclusive semileptonic
decays $\Lambda_b\to \Lambda_c e^- \bar\nu_e$ and
$\Lambda_c\to \Lambda e^+ \nu_e$ for the finite and infinite mass cases.
In Table I we present our results for the two sets of relativistic
form factors (Set I and Set II)\footnote{ We use the notation:\\
Set I - the set of form factors $(F_1^V, F_2^V, F_3^V, F_1^A, F_2^A, F_3^A)$,
\\
Set II - the set of form factors
$(G_1^V, G_2^V, G_3^V, G_1^A, G_2^A, G_3^A).$}
for two values of $q^2$, namely $q^2=q^2_{max}=(M_1-M_2)^2$
and $q^2=0$. In parenthesis we give the values of the form factors
in the  heavy quark limit.
Table I shows that in the case of the $b\to c$ transition
the ``finite mass'' corrections can amount to $\sim 10\%$
for the Set I form factors and up to $\sim 25\%$ for the Set II form
factors. In the case of
$c\to s$ transitions the ``finite mass'' corrections are significantly
larger (see TABLE II). Our estimate of  the ``finite mass'' corrections for
$\Lambda_b\to\Lambda_c$ transition at $q^2=q^2_{max}$
agrees with the QCD sum rules calculations:
$F_1^V=F_1^A=1.03\pm 0.06$ and $F_2^{V(A)}=F_3^{V(A)}=0$ \cite{DFNR}.
In TABLE III we show our results for the rates and mean asymmetries of  
$\Lambda_b$ and $\Lambda_c$ decays again for the finite and infinite
mass cases. For the heavy-to-heavy transition
$\Lambda_b\to \Lambda_c e^- \bar\nu_e$ the finite mass corrections are
$\sim 10\%$ in the decay rate and $\sim 7\%$ in the asymmetry parameter.
In the case of $\Lambda_c\to \Lambda_s$ transition one can see that
the finite mass corrections are larger than in the 
$\Lambda_b\to \Lambda_c$ case.
They amount to $\sim 50\%$ for the decay rate
and $\sim 10\%$ for the asymmetry parameter. Our prediction for the
$\Lambda_b\to\Lambda_c e^- \bar\nu_e$ rate is in agreement with the
experimental upper limit given by
$\Gamma(\Lambda_b\to\Lambda_c e^- \bar\nu_e)= (6.67\pm 2.73)\times 10^{10}$
sec$^{-1}$ \cite{PDG}. Our prediction for the
$\Lambda_c\to\Lambda_s e^+ \nu_e$ rate agrees with
the the corresponding experimental value measured by the CLEO Collaboration:
$\Gamma(\Lambda_c\to\Lambda_s e^+ \nu_e)= (9.54\pm 2.28)\times 10^{10}$
sec$^{-1}$ \cite{PDG}.

In conclusion, we have found that the finite mass corrections
significantly contribute to the rate and the asymmetry parameter
in the heavy-to-light transition $\Lambda_c\to\Lambda_s e^+ \nu_e$
whereas the finite mass corrections are smaller for the heavy-to-heavy
transition $\Lambda_b\to\Lambda_c e^- \bar\nu_e$. It appears that the
leading term in the heavy quark expansion gives a reasonably accurate
description of the heavy-to-heavy $b\to c$ transitions.
Contrary to this the finite mass corrections are quite important for
the heavy-to-light $c\to s$ transitions.

M.A.I. and V.E.L. thank Mainz University for hospitality where  
this work was completed. The visit of M.A.I. to Mainz University 
was supported by the DFG (Germany) and the visit of V.E.L. to Mainz 
University was supported by the Graduiertenkolleg "Eichtheorien'', Mainz.  
This work was supported in part by the Heisenberg-Landau Program
and by the BMBF (Germany) under contract 06MZ865. J.G.K. acknowledges
partial support by the BMBF (Germany) under contract 06MZ865.
A.G.R. acknowledges partial support of the Swiss National Science
Foundation, and TMR, BBW-Contract No. 97.0131  and  EC-Contract
No. ERBFMRX-CT980169 (EURODA$\Phi$NE).

\vspace*{1cm}
%%%%%%%%%%%%%%%%%%%%%%%%%%%%%%%%%%%%%%%%%%%%%%%%%%%%%%%%
%                      TABLES
%%%%%%%%%%%%%%%%%%%%%%%%%%%%%%%%%%%%%%%%%%%%%%%%%%%%%%%
%

\centerline{\bf List of Tables}

{\bf TABLE I}
Form Factors of $\Lambda_b\to\Lambda_c$ transitions.

\vspace*{.5cm}
{\bf TABLE II}
Form Factors of $\Lambda_c\to\Lambda_s$ transitions.

\vspace*{.5cm}
{\bf TABLE III}
Exclusive decay characteristics of $\Lambda_b$ and $\Lambda_c$ baryons.

\newpage
\begin{center}
{\bf TABLE I}
\end{center}
\small \normalsize
\begin{tabular}{|c|c|c|c|c|c|c|}
\hline
  Set  &\multicolumn{6}{|c|}{$q^2=q^2_{max}$} \\
       \cline{1-7}
    I  &$F_1^V=0.99$ & $F_2^V=0.034$ &$F_3^V=0.001$
       &$F_1^A=0.97$ & $F_2^A=-0.002$&$F_3^A=-0.035$\\
       &$(1)$ & $(0)$ & $(0)$ & $(1)$ & $(0)$ & $(0)$ \\
   II  &$G_1^V=1.26$ & $G_2^V=-0.20$ &$G_3^V=-0.067$
       &$G_1^A=0.96$ & $G_2^A=-0.21$ &$G_3^A=-0.076$\\
       & $(1)$ & $(0)$ & $(0)$ & $(1)$ & $(0)$ & $(0)$ \\
\hline
  Set  &\multicolumn{6}{|c|}{$q^2=0$} \\
       \cline{1-7}
    I  &$F_1^V=0.55$ & $F_2^V=0.017$ &$F_3^V=0.005$
       &$F_1^A=0.54$ & $F_2^A=-0.001$ &$F_3^A=-0.017$ \\
       & $(0.62)$ & $(0)$ & $(0)$ & $(0.62)$ & $(0)$ & $(0)$ \\
   II  &$G_1^V=0.69$ & $G_2^V=-0.10$ &$G_3^V=-0.033$
       &$G_1^A=0.54$ & $G_2^A=-0.10$ &$G_3^A=-0.036$ \\
       & $(0.62)$ & $(0)$ & $(0)$ & $(0.62)$ & $(0)$ & $(0)$ \\
\hline
\end{tabular}

\vspace*{.2cm}
\begin{center}
{\bf TABLE II}
\end{center}
\small \normalsize
\begin{tabular}{|c|c|c|c|c|c|c|}
\hline
 Set   & \multicolumn{6}{|c|}{$q^2=q^2_{max}$} \\
       \cline{1-7}
   I   &$F_1^V=0.70$ & $F_2^V=-0.14$ &$F_3^V=-0.045$
       &$F_1^A=0.76$ & $F_2^A=-0.032$ &$F_3^A=-0.16$\\
       & $(0.62)$ & $(-0.044)$ & $(-0.044)$
       & $(0.63)$ & $(-0.044)$ & $(-0.044)$ \\
  II   &$G_1^V=1.16$ & $G_2^V=-0.41$ &$G_3^V=-0.10$
       &$G_1^A=0.65$ & $G_2^A=-0.45$ &$G_3^A=0.15$\\
       & $(0.77)$ & $(-0.20)$ & $(-0.20)$
       & $(0.58)$ & $(-0.20)$ & $(-0.20)$\\
\hline
 Set   & \multicolumn{6}{|c|}{$q^2=0$} \\
       \cline{1-7}
    I  &$F_1^V=0.41$ & $F_2^V=-0.071$ &$F_3^V=-0.025$
       &$F_1^A=0.38$ & $F_2^A=-0.021$ &$F_3^A=-0.084$ \\
       & $(0.33)$ & $(-0.018)$ & $(-0.018)$
       & $(0.33)$ & $(-0.018)$ & $(-0.018)$ \\
   II  &$G_1^V=0.65$ & $G_2^V=-0.22$ &$G_3^V=-0.050$
       &$G_1^A=0.34$ & $G_2^A=-0.24$ &$G_3^A=0.070$ \\
       &$(0.39)$ & $(-0.08)$ & $(0)$ & $(0.31)$ & $(-0.08)$ & $(0)$ \\
\hline
\end{tabular}

\vspace*{.5cm}

\begin{center}
{\bf TABLE III}
\end{center}
\small \normalsize
\begin{center}
\begin{tabular}{|c|c|c|c|c|c|c|}
\hline
Process &\multicolumn{2}{|c|}{Heavy Quark Limit} &
        \multicolumn{2}{|c|}{Finite Quark Masses} \\
       \cline{1-5}
$\Lambda_b\to\Lambda_c+e^-\bar\nu_e$ & $\Gamma=5.4\times 10^{10}sec^{-1}$
                                     & $\alpha=-0.761$
                                     & $\Gamma=4.9\times 10^{10}sec^{-1}$
                                     & $\alpha=-0.815$\\
\hline
$\Lambda_c\to\Lambda_s+e^+\nu_e$ & $\Gamma=11.8\times 10^{10}sec^{-1}$
                                 & $\alpha=-0.798$
                                 & $\Gamma=7.6\times 10^{10}sec^{-1}$
                                 & $\alpha=-0.878$\\
 \hline
\end{tabular}
\end{center}

%%%%%%%%%%%%%%%%%%%%%%%%%%%%%%%%%%%%%%%%%%%%%%%%%
%            FIGURES
%%%%%%%%%%%%%%%%%%%%%%%%%%%%%%%%%%%%%%%%%%%%%%%%%%
%

\newpage
\centerline{\bf List of Figures}

{\bf Fig. 1} \hspace*{.2cm} Semileptonic decay of heavy baryon
$\Lambda_{b(c)}\to\Lambda_{c(s)} + e +\nu_e$.

\vspace*{1.5cm}
\unitlength=1.00mm
\special{em:linewidth 0.4pt}
\linethickness{0.4pt}
\begin{picture}(115.00,64.00)
\put(50.00,20.00){\circle*{5.20}}
\put(95.00,20.00){\circle*{5.20}}
\put(73.00,42.00){\circle*{3.00}}
\put(73.00,5.00){\makebox(0,0)[cc]{\Large$\bf d$}}
\put(30.00,26.00){\makebox(0,0)[cc]{\Large$\bf \Lambda_{b (c)}$}}
\put(115.00,26.00){\makebox(0,0)[cc]{\Large$\bf \Lambda_{c (s)}$}}
\put(73.00,23.20){\makebox(0,0)[cc]{\Large$\bf u$}}
\put(56.00,34.20){\makebox(0,0)[cc]{\Large$\bf b (c)$}}
\put(90.00,34.20){\makebox(0,0)[cc]{\Large$\bf c (s)$}}
\put(72.50,17.00){\oval(45.00,16.00)[b]}
\put(50.00,20.00){\line(1,1){22.00}}
\put(30.00,20.00){\line(1,0){85.00}}
\put(30.00,19.00){\line(1,0){20.00}}
\put(30.00,21.00){\line(1,0){20.00}}
\put(95.00,20.00){\line(0,-1){1.00}}
\put(95.00,19.00){\line(1,0){20.00}}
\put(95.00,21.00){\line(1,0){20.00}}
\put(95.00,20.00){\line(-1,1){22.00}}
\put(72.85,42.00){\line(0,1){2.00}}
\put(72.85,45.00){\line(0,1){2.00}}
\put(72.85,48.00){\line(0,1){2.00}}
\put(72.85,51.00){\line(0,1){2.00}}
\put(72.85,54.00){\line(0,1){2.00}}
\put(72.85,57.00){\line(0,1){2.00}}
\put(72.85,60.00){\line(0,1){2.00}}
\put(72.85,63.00){\line(0,1){2.00}}
\put(75,69){\makebox(0,0)[cc]{\LARGE$e\nu_e$}}
\end{picture}

\newpage

\noindent
\end{document}